\documentclass[twocolumn,letterpaper,amsmath,amssymb,floatfix,aps,superscriptaddress]{revtex4}

\usepackage{graphicx}
\usepackage{dcolumn}
\usepackage{bm}
\usepackage[usenames]{color}
\usepackage{hyperref}
\usepackage{ulem}

\begin{document}

\title{Pulling a folded polymer through a nanopore}

\author{Bappa Ghosh}
\thanks{These two authors contributed equally}
\affiliation{Department of Chemistry, Indian Institute of Science Education and Research, 
Pune, Maharashtra, India.}

\author{Jalal Sarabadani}
\thanks{These two authors contributed equally}
\email{jalal@ipm.ir}
\affiliation{School of Nano Science, Institute for Research in Fundamental Sciences (IPM), 19395-5531, Tehran, Iran}

\author{Srabanti Chaudhury}
\email{srabanti@iiserpune.ac.in}
\affiliation{Department of Chemistry, Indian Institute of Science Education and Research, 
Pune, Maharashtra, India.}

\author{Tapio Ala-Nissila}
\affiliation{Department of Applied Physics and QTF Center of Excellence, Aalto University 
School of Science, P.O. Box 11000, FI-00076 Aalto, Espoo, Finland.}
\affiliation{Interdisciplinary Centre for Mathematical Modelling and Department of Mathematical Sciences, 
Loughborough University, Loughborough, Leicestershire LE11 3TU, UK.}

\begin{abstract}
We investigate the translocation dynamics of a folded linear polymer which is pulled through a nanopore by an external force.
To this end, we generalize the iso-flux tension propagation
(IFTP) theory for end-pulled polymer translocation to include the case of two segments of the folded polymer traversing simultaneously trough the pore.
Our theory is extensively benchmarked with corresponding Molecular Dynamics (MD) simulations.
The translocation process for a folded polymer can be divided into two main stages. In the first stage, both branches are 
traversing the pore and their dynamics is coupled. If the 
branches are not of equal length, there is a second stage where translocation of the shorter branch has been completed.
Using the assumption of equal monomer flux of both branches, we analytically derive the equations of motion for both branches and
characterise the translocation dynamics in detail
from the average waiting time and its scaling form. Moreover, MD simulations are used to study
additional details of translocation dynamics such as the translocation time distribution and individual monomer velocities.
\end{abstract}

\maketitle


The process of polymer translocation through nanometer sized pores plays an important role 
in many biological \cite{Akeson1999,Lingappa1984} as well as technological applications \cite{Turner2002,Chang_book}.
Experiments using single molecule precision have stimulated many theoretical and
computational studies on polymer translocation 
\cite{kasi1996,mellerPRL2001,sung1996,muthu1999,muthu2003,sakaue2007,sakaue2010,saito2012,ikonen2012a,%
ikonen2012b,ikonen2013,ikonen2012c,Tapio_review,%
jalal2014,jalal2015,JalalEPL2017,jalalSciRep2017,jalalJPC2018,unbiased_Slater_1,unbiased_Slater_2,%
unbiased_Slater_3,golestanianPRL2011,golestanianPRX2012,golestanianJCP2012,tapioPRL2008DNAsequencing,%
aksimentievNanolett2008,Milchev_JPCM}. Over the last few years a comprehensive theory of driven translocation dynamics 
has been developed based on the idea of tension propagation in the chain. This iso-flux tension propagation (IFTP) theory has been
applied to a variety of different physical scenarios \cite{jalalJPC2018}, including 
pore-driven translocation of flexible \cite{jalal2014}
and semi-flexible \cite{jalalSciRep2017} polymers, end-pulled \cite{JalalEPL2017} polymers, and translocation of a flexible polymer through 
a flickering nanopore 
under an alternating external driving force acting in the pore \cite{jalal2015}.

\begin{figure*}[t]
   \begin{center}
        \includegraphics[width=0.999\linewidth]{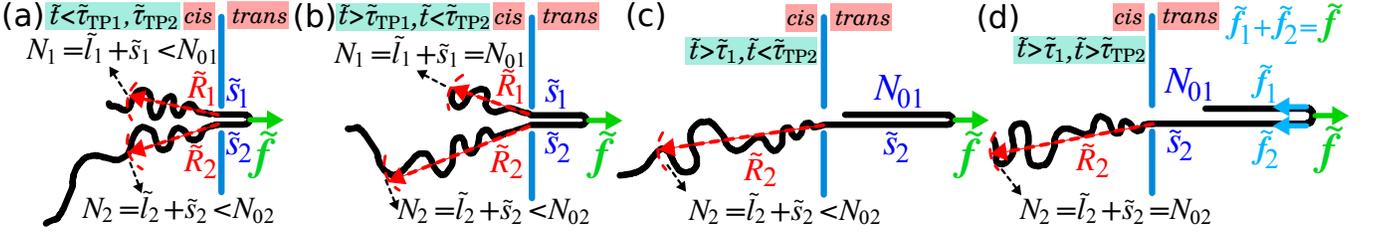}
    \end{center}
\caption{(a) Schematic of the translocation process of a polymer with a folded configuration. 
The contour lengths for shorter and the longer branches are $N_{01}$ and $N_{02}$, respectively. 
The external driving force, $\tilde{f}$, acts on the pulled monomer that connects 
the shorter branch to the longer one. The direction of the force is from the {\it cis} towards the {\it trans} side. 
For both branches the translocation process is in the tension propagation (TP) stage, where the tension forces have 
not reached the branches ends and $N_{\textrm{b}} = \tilde{l}_{\textrm{b}} + \tilde{s}_{\textrm{b}} < N_{0{\textrm{b}}}$ 
with ${\textrm{b}}=1$ and 2 denoting the branch ${\textrm{b}}$ (TP1-TP2).
For each branch the location of the tension front divides the mobile part from the immobile equilibrium part, 
and is represented by $\tilde{R}_1$ and $\tilde{R}_2$ for the shorter and longer branches, respectively.
The number of translocated monomers (translocation coordinates) for shorter and longer branches 
are $\tilde{s}_1$ and $\tilde{s}_2$, respectively.
(b) The translocation process is in TP stage for the longer branch ($N_2 = \tilde{l}_2 + \tilde{s}_2 < N_{02}$), 
while it is in post propagation (PP) stage for shorter one and $N_1 = \tilde{l}_1 + \tilde{s}_1 = N_{01}$,
where the tension force has already reached the shorter branch end (PP1-TP2).
(c) The translocation process for the shorter branch has been completed while it is still in TP stage for the longer one
($\tau_1$-TP2 stage, where $\tau_1$ stands for the translocation time of the shorter branch).
(d) The same as panel (c) but the translocation process is in PP stage for the longer branch ($\tau_1$-PP2).
Here, the contribution of the total external driving force, $\tilde{f}$, to each branch has been shown as 
$\tilde{f}_1$ and $\tilde{f}_2$, and force balance for the pulled monomer gives 
$\tilde{f}_1 + \tilde{f}_2 = \tilde{f}$ (not shown in panels (a)-(c)).
} 
\label{fig_schematic}
\end{figure*}

Experiments on polymer 
translocation typically involve the electrically driven movement of charged polymers through pores whose typical diameters 
range from nanometers to tens of nanometers. In most of these experiments, the $\alpha$-hemolysin pore complex is 
used in common to study the translocation process \cite{Akeson1999,kasi1996,Mathe2005}. 
These experimental studies have been effective in distinguishing polymers of different molecular weights and sequences.  
However, there are several limitations in the use of this approach. The $\alpha$-hemolysin pore has a diameter about 
2 nm such that only single-stranded DNA/RNA molecules or synthetic polyelectrolytes are restricted to thread 
through these protein channels. In the experiment by Kasionawicz {\it et al.}, the ionic current through the voltage 
biased $\alpha$-hemolysin pore can detect the translocation of single-stranded molecules through the narrow pore 
under the influence of an external field \cite{kasi1996}. In addition, $\alpha$-hemolysin is not stable at wide 
experimental conditions such as high voltages ranges, temperatures, pH etc.

To overcome these difficulties, artificial 
solid-state nanopores have been developed and applied for studying polymer translocation. These nanopores can be tuned 
to larger diameters of $10-20$ nm that allows translocation of double-stranded DNA molecules. Experiments on double-strand DNA 
translocation using silicon oxide nanopores have been reported by Dekker and coworkers 
\cite{Dekker2007, Storm2003, Keyser2006, Heng2005}. 
Such synthetic pores have a lot of advantages over biological pores. For example, the synthetic pores are stable under experimental 
conditions such as high temperature, extreme voltage and pH values 
\cite{Nakane2002, Kasianowicz2001, Bayley2001, Bayley2000, storm2005}. 
Since solid-state pores can have larger diameters, it has been observed experimentally that a polymer can undergo not 
only single file motion through the pore but also in different folded states \cite{Storm2005_2}. The formation of double-stranded 
DNA hairpins undergoing voltage-driven translocation through nanopores located in synthetic membranes has been studied using 
coarse-grained Langevin dynamics of translocation \cite{Forrey2007}. Kotsev and Kolomeisky have given a theoretical description 
of the translocation dynamics of polymer with folded configurations using simple discrete stochastic models \cite{Kotsev2007}. 
The translocation dynamics is considered as the motion of the folded segment of the chain through the channel followed by 
the motion of the linear part of the polymer. However, driven polymer translocation is controlled by tension front propagation and a
proper theoretical treatment using the IFTP theory
has not been done to date.

To this end, here we study the translocation dynamics of a pulled folded polymer through a nanopore \cite{JalalEPL2017}
by generalizing the IFTP theory to include the simultaneous translocation of two polymer strands in the pore. The theory
is benchmarked with MD simulations.
We consider both the symmetric case where the polymer is pulled in the middle monomer such that branches have equal lengths, and the asymmetric case with
unequal branch lengths, as shown in Fig.~\ref{fig_schematic}. 
The details of the modified IFTP theory and MD simulations can be found in Secs.~\ref{model} and \ref{MD}, 
respectively, and the results are discussed in Sec.~\ref{results}. Finally, Sec.~\ref{conclusions} is
devoted to present the summary and conclusions.


\section{Theoretical model} \label{model}

In this section we generalize the IFTP theory for end-pulled translocation dynamics to the present case of a folded linear polymer.  
When the contour lengths of both branches are the same, they traverse through the pore at equal rates on average. 
In contrast when the contour length of branches are not equal 
the translocation time of the shorter branch is less than that of the longer one.
To develop the IFTP theory we consider here the high force limit and assume that the {\it trans} side sub-branches 
are fully straightened. This imposes the condition that the monomer flux which is the number of monomers that pass
through the pore per unit time is the same for both branches. 
The dynamics of each branch is studied separately in the presence of the other one which leads to coupling between 
their equations of motion.

For the sake of simplicity, in the IFTP theory dimensionless units denoted by tilde are used
as $\tilde{Y} \equiv Y / Y_u$, with the units of time $t_u \equiv \eta \sigma^2 / (k_{\rm B} T)$,
length $s_u \equiv \sigma$, velocity $v_u \equiv \sigma/t_u = k_{\rm B} T/(\eta \sigma)$, 
force $f_u \equiv k_{\rm B} T/\sigma$, monomer flux $\phi_u \equiv k_{\rm B} T/(\eta \sigma^2)$ 
and friction $\Gamma_u \equiv \eta$, where $T$ is the temperature of the system, $k_{\rm B}$ 
is the Boltzmann constant, $\sigma$ is the length of each segment, and the solvent friction 
per monomer is $\eta$.
The quantities without the tilde are expressed in the Lennard-Jones units.

In Fig.~\ref{fig_schematic} we show a schematic of the translocation process of a folded polymer. 
The contour lengths for short and the long branches are denoted by $N_{01}$ and $N_{02}$, respectively, 
i.e. $N_{01} \leq N_{02}$.
The external driving force, $\tilde{f}$, acts on the monomer which connects 
the short branch to the long one, and its direction is from {\it cis} towards the {\it trans} side. 
Panel (a) illustrates the tension propagation (TP) stage for both branches (TP1-TP2), wherein the 
tension force has not reached the branches' ends and 
$N_{\textrm{b}} = \tilde{l}_{\textrm{b}} + \tilde{s}_{\textrm{b}} < N_{0{\textrm{b}}}$ 
(${\textrm{b}}=1$ and 2 stand for short and long branches, respectively). 
$N_{\textrm{b}}$ is the total number of beads in branch $\textrm{b}$ that have been already affected by the tension force,
and $\tilde{l}_{\textrm{b}}$ is the number of beads in the mobile domain of the {\it cis} side sub-branch b.
In the TP stage the location of the tension front separates the mobile sub-branches from the immobile equilibrium ones, 
and is represented by $\tilde{R}_{\textrm{b}}$. During the TP stage it is assumed that $\tilde{R}_1 = \tilde{R}_2$.
The number of translocated monomers (translocation coordinates) for shorter and longer branches 
are $\tilde{s}_1$ and $\tilde{s}_2$, respectively. 
When both branches are inside the pore then $\tilde{s}_1 = \tilde{s}_2$. This is due to the strong force limit,
wherein both sub-branches in the {\it trans} side are straightened.
As time passes the shorter branch experiences the post propagation (PP1) stage,
where the tension force has reached its end, and $N_1 = \tilde{l}_1 + \tilde{s}_1 = N_{01}$,
while the longer one is still in the TP stage (PP1-TP2). This is illustrated in panel (b).
Then as can be seen in panel (c), the translocation process for the short branch has been completed 
while it is still in TP stage for the long one ($\tau_1$-TP2, where $\tau_1$ is the translocation 
time for the shorter branch). Finally, panel (d) shows that the translocation process 
is in PP stage for the long branch ($\tau_1$-PP2). Moreover, in panel (d), 
the contribution of the total external driving force, $\tilde{f}$, to each branch is illustrated
as $\tilde{f}_1$ and $\tilde{f}_2$. The force balance for the pulled monomer gives 
$\tilde{f}_1 + \tilde{f}_2 = \tilde{f}$ (not shown in panels (a)-(c)).

In addition, in the very strong force limit of $\tilde{f}_{0\textrm{b}} > \tilde{N}_{\textrm{b}}$ 
the mobile sub-branch b in the {\it cis} side is fully straightened (strong stretching (SSC) limit). 
In the moderate external force limit of $1 < \tilde{f}_{0\textrm{b}} < \tilde{N}_{\textrm{b}}$ 
the regime is called stem-flower (SFC) as the shape of the mobile sub-branch b in the {\it cis} side 
is similar to a stem followed by a flower. Finally, the weak force limit of $\tilde{f}_{0\textrm{b}} < 1$ 
is called the trumpet (TRC) regime, where the mobile sub-branch in the {\it cis} side assumes a trumpet shape.

As mentioned above and depicted in Fig.~\ref{fig_schematic} only the SS regime in the {\it trans} side 
is considered here, and therefore using the deterministic version of the iso-flux Brownian dynamics tension 
propagation theory without any entropic force~\cite{ikonen2012a,jalal2014} is a very good approximation. 
Within this framework, the equation of motion for the time evolution of the translocation coordinate 
$\tilde{s}_{\textrm{b}}$ for branch $\textrm{b}$, that is the number of translocated beads to the 
{\it trans} side, is written as
\begin{equation}
\tilde{\Gamma}_{\textrm{b}} (\tilde{t}) \frac{ \textrm{d} \tilde{s}_{\textrm{b}} }{ \textrm{d} \tilde{t} } = 
\tilde{f}_{\textrm{b}} ,
\label{BD_force}
\end{equation}
where $\tilde{\Gamma}_{\textrm{b}} (\tilde{t})$ is the effective friction and $\tilde{f}_{\textrm{b}}$
is the contribution of the total external driving force $\tilde{f}$ to branch $\textrm{b}$ as depicted 
in Fig.~\ref{fig_schematic}(d). According to Refs. \cite{jalalSciRep2017,JalalEPL2017} the effective 
friction can be obtained as 
$\tilde{\Gamma}_{\textrm{b}} (\tilde{t}) = \tilde{\eta}_{{cis}\textrm{b}} (\tilde{t})
+ \tilde{\eta}_{\textrm{pj}} (\tilde{t}) + \tilde{\eta}_{\textrm{TSb}} (\tilde{t}) $,
where $\tilde{\eta}_{{cis}\textrm{b}} (\tilde{t})$ denotes the friction due the {\it cis} side mobile 
sub-branch $\textrm{b}$ in the solvent, $\tilde{\eta}_{\textrm{pj}} (\tilde{t})$ is the pore friction, 
and $\tilde{\eta}_{\textrm{TSb}} (\tilde{t}) $ presents the friction due the movement of the 
{\it trans} side mobile sub-branch $\textrm{b}$.
When both branches are inside the pore, i.e. in the stages TP1-TP2, or PP1-TP2, 
$\textrm{j} = 12$ and when only the longer one is inside the pore, i.e. in the stages $\tau_1$-TP2 or $\tau_1$-PP2,
$\textrm{j} = 2$.
It should be mentioned that the {\it trans} side friction terms play an important role 
in the dynamics of the current system \cite{jalalSciRep2017,JalalEPL2017}. This will be discussed in detail below.
Later it will be shown how to find the values of $\tilde{f}_1$ and $\tilde{f}_2$ during the translocation process. 
In the symmetric case when the contour lengths of both branches are the same, i.e. $N_{01} = N_{02}$, 
then due to the symmetry of the system $\tilde{f}_1 = \tilde{f}_2 = \tilde{f} /2$ during the whole translocation process.

Using the IFTP theory the dynamics of each branch in the {\it cis} and in the {\it trans} sides is separately
solved with the corresponding TP equations. 
The iso-flux (IF) approximation is used to find the TP equations \cite{rowghanian2011}. In the IF approximation
the monomer flux $\tilde{\phi} (\tilde{t}) = \textrm{d} \tilde{s} / \textrm{d} \tilde{t} $ within the mobile domain 
for each branch is constant in space but evolves with time. 
In the TP1-TP2 and PP1-TP2 stages the tension front is located at distance 
$\tilde{x} = \tilde{R}_{\textrm{b}} (\tilde{t})$ to the pore in the {\it cis} side. 
Inside each branch, the tension force is mediated from the pulled monomer at the distance $\tilde{s}$
in the {\it trans} side all the way to the pore located at $\tilde{x} = 0$ and then to the last mobile
bead $N_{\textrm{b}}$ located in the tension front in the {\it cis} side. 
Performing the integration of the local force-balance relation 
$\textrm{d} \tilde{f}_{\textrm{b}} (\tilde{x}') = -\tilde{\phi}_{\textrm{b}} (\tilde{t}) \textrm{d} \tilde{x}' $ 
\cite{jalal2014,JalalEPL2017}
over the distance from the location of the pulled monomer to $\tilde{x}$, gives the 
tension force at the distance $\tilde{x}$ as
\begin{equation}
\tilde{f}_{\textrm{b}} (\tilde{x} , \tilde{t}) =  
\tilde{f}_{0{\textrm{b}}}
- \tilde{x} \tilde{\phi}_{\textrm{b}} (\tilde{t}),
\label{tension_force}
\end{equation}
where $\tilde{f}_{0{\textrm{b}}} = \tilde{f}_{\textrm{b}} - \tilde{\eta}_{\textrm{pj}} \tilde{\phi}_{\textrm{b}} (\tilde{t}) 
- \tilde{\eta}_{\textrm{TSb}} \tilde{\phi}_{\textrm{b}} (\tilde{t}) $
is the force at the entrance of the pore in the {\it cis} side.
Combining Eq.~(\ref{tension_force}) and the fact that the tension force vanishes at the tension front, i.e. 
$\tilde{f}_{\textrm{b}} (\tilde{R}_{\textrm{b}} , \tilde{t}) =  0$ yields the monomer flux as 
\begin{equation}
\tilde{\phi}_{\textrm{b}} (\tilde{t}) =
\frac{ \tilde{f}_{\textrm{b}} }{ \tilde{R}_{\textrm{b}} + \tilde{\eta}_{\textrm{pj}} + \tilde{\eta}_{\textrm{TSb}} }.
\label{monomer_flux}
\end{equation}
Since we are in the SS limit for both sub-branches in the {\it trans} side, 
$\tilde{\eta}_{\textrm{TSb}} = \tilde{s}_{\textrm{b}} = \tilde{s}$.
Moreover, when both branches are inside the pore $\tilde{\eta}_{\textrm{pj}}= \tilde{\eta}_{\textrm{p12}} $,
and if only the long branch is located in the pore $\tilde{\eta}_{\textrm{pj}}= \tilde{\eta}_{\textrm{p2}} $ .

To determine $\tilde{f}_1$ and $\tilde{f}_2$, two equations must be solved.
The first equation is $\tilde{f}_1 + \tilde{f}_2 = \tilde{f}$, which is the force balance equation for the 
pulled monomer as shown in Fig.~\ref{fig_schematic}(d). The second one comes from the fact that 
monomer fluxes for both branches are the same, i.e. $\tilde{\phi}_{\textrm{1}} (\tilde{t}) = \tilde{\phi}_{\textrm{2}} (\tilde{t})$,
where $\tilde{\phi}_{\textrm{b}} (\tilde{t}) $ has been defined in Eq.~(\ref{monomer_flux}).
Solving these two equations gives
\begin{eqnarray}
\tilde{f}_{\textrm{1}} &=&  \tilde{f} \times \bigg( 
1 + \frac{ \tilde{R}_2 + \tilde{\eta}_{ \textrm{TS2}} + \tilde{\eta}_{ \textrm{p12}} }
{ \tilde{R}_1 + \tilde{\eta}_{ \textrm{TS1}} + \tilde{\eta}_{ \textrm{p12}} }  \bigg)^{-1}  ; \nonumber\\
\tilde{f}_{\textrm{2}} &=&  \tilde{f} \times \bigg( 
1 + \frac{ \tilde{R}_1 + \tilde{\eta}_{ \textrm{TS1}} + \tilde{\eta}_{ \textrm{p12}} }
{ \tilde{R}_2 + \tilde{\eta}_{ \textrm{TS2}} + \tilde{\eta}_{ \textrm{p12}} }  \bigg)^{-1}  .
\label{force_1_2}
\end{eqnarray}
Inserting the above $\tilde{f}_{\textrm{b}}$ into Eq.~(\ref{monomer_flux}), the monomer flux reads as
\begin{equation}
\tilde{\phi}_1 (\tilde{t}) = \tilde{\phi}_2 (\tilde{t}) = \tilde{\phi} (\tilde{t}) =
\frac{ \tilde{f} }{ \tilde{R}_1 + \tilde{R}_2 + 2 \tilde{\eta}_{\textrm{p12}} + \tilde{\eta}_{\textrm{TS1}} 
+ \tilde{\eta}_{\textrm{TS2}} }.
\label{monomer_flux_12}
\end{equation}
If the contour lengths for both branches are the same, or the translocation process is in 
TP1-TP2, then due to the symmetry $\tilde{R}_1 = \tilde{R}_2$.
This leads to $\tilde{f}_1 = \tilde{f}_2  = \tilde{f} /2$, and consequently 
$\tilde{\phi} (\tilde{t}) = (1/2) \tilde{f} \times 
\big( \tilde{R} + \tilde{\eta}_{\textrm{p12}} + \tilde{s} \big)^{-1}$.

In the $\tau_1$-TP2 and $\tau_1$-PP2 stages where the whole short branch has been translocated
to the {\it trans} side, the time evolution of $\tilde{s}_2$ is given by Eq.~(\ref{BD_force}) with 
index $\textrm{b}=2$, and similar procedure to the TP1-TP2 and PP1-TP2 stages is employed to obtain 
the monomer flux as
\begin{equation}
\tilde{\phi} (\tilde{t}) =
\frac{ \tilde{f} }{ \tilde{R}_2 + \tilde{\eta}_{\textrm{p2}} + \tilde{\eta}_{\textrm{TS2}} + N_{01} },
\label{monomer_flux_22}
\end{equation}
where $\tilde{\eta}_{\textrm{p2}}$ is the pore friction when only the long branch is inside the pore,
$ \tilde{\eta}_{\textrm{TS2}} = \tilde{s}_2$, and $N_{01}$ is the {\it trans} side friction 
due to the whole mobile short branch.

In the TP1-TP2 and PP1-TP2 stages combining Eqs.~(\ref{BD_force}) and (\ref{monomer_flux}) 
and (\ref{force_1_2}), the time evolution of the translocation coordinates are obtained for 
short and long branches provided that the time evolution of the location of the tension 
front for each branch is known. 
On the other hand in the $\tau_1$-TP2 and $\tau_1$-PP2 stages 
Eq.~(\ref{BD_force}) together with (\ref{monomer_flux_22}) give the translocation coordinate
for the longer branch as a function of time again if the location of the tension front for
the long branch is know.
Therefore, to proceed further the time evolution of the location of the tension fronts
for the short as well as long branches, $\tilde{R}_1$ and $\tilde{R}_2$, should be obtained. 

To obtain the time evolution of $\tilde{R}_{\textrm{b}}$ in the TP stage one can use the 
end-to-end distance $\tilde{R}_{\textrm{b}} = A_{\nu} N_{\textrm{b}}^{\nu}$, where $A_{\nu} = 1.15$
is constant obtained from MD simulations, $N_{\textrm{b}} = \tilde{l}_{\textrm{b}} + \tilde{s}_{\textrm{b}}$,
and $\nu=0.5888$ is the Flory exponent for 3D. 
The time derivative of the above relation is then written as 
$\dot{\tilde{R}}_{\textrm{b}} = A_{\nu}^{1/\nu} \tilde{R}_{\textrm{b}}^{(\nu-1)/\nu} 
\big( \dot{\tilde{l}}_{\textrm{b}} + \dot{\tilde{s}}_{\textrm{b}}\big) $, where $\dot{\tilde{s}}_{\textrm{b}} $
is the monomer flux and $\dot{\tilde{l}}_{\textrm{b}}$ must be found. 
In the SSC regime as the mobile sub-branch b in the {\it cis} side is fully straightened the 
monomer number density is unity, 
while according to the blob theory in the SFC and TRC regimes the monomer number density, which is
$\tilde{\sigma}_{\textrm{b}} (\tilde{x}) = |\tilde{f}_{\textrm{b}} (\tilde{x})|^{(\nu-1)/\nu}$,
is larger than unity due to the folding of the chain.
By integrating $\tilde{\sigma}_{\textrm{b}}^{\textrm{I}} ( \tilde{x})$ ($\textrm{I}=$SSC, SFC or TRC) 
over the distance from the pore entrance in the {\it cis} side to the location of the tension front,
$\tilde{l}_{\textrm{b}}^{\textrm{I}}$ is obtained as
\begin{eqnarray}
\tilde{l}_{\textrm{b}}^{\textrm{SSC}} &=& 
\tilde{R}_{\textrm{b}}; \nonumber\\
\tilde{l}_{\textrm{b}}^{\textrm{SFC}} &=& 
\tilde{R}_{\textrm{b}} + \frac{1-\nu}{2\nu-1} \frac{1}{\tilde{\phi}_{\textrm{b}}} ; \nonumber\\
\tilde{l}_{\textrm{b}}^{\textrm{TRC}} &=& 
\frac{\nu}{2\nu-1}  \tilde{R}_{\textrm{b}}^{(2\nu-1)/\nu} \tilde{\phi}_{\textrm{b}}^{(\nu-1)/\nu} .
\label{el}
\end{eqnarray}
Combining the time derivative of $\tilde{l}_{\textrm{b}}^{\textrm{I}}$ with the above relation for 
$\dot{\tilde{R}}_{\textrm{b}}$, the equations of motion for the location of the tension front
when both branches are inside the pore are obtained as 
\begin{equation}
\dot{\tilde{R}}_{\textrm{1}}^{\textrm{I,J},12} = \mathcal{U}_{\textrm{1}}^{\textrm{I,J}} ; \hspace{+1.0cm}
\dot{\tilde{R}}_{\textrm{2}}^{\textrm{I,J},12} = \mathcal{U}_{\textrm{2}}^{\textrm{I,J}} ,
\label{R_12}
\end{equation}
and when only the longer branch remains inside the pore
\begin{equation}
\dot{\tilde{R}}_{\textrm{2}}^{\textrm{I,J},2} = \mathcal{V}_{\textrm{2}}^{\textrm{I,J}} ,
\label{R_22}
\end{equation}
where $\textrm{I}$ denotes the different regimes of SSC, SFC and TRC, $\textrm{J}$ stands for different stages of TP and PP, 
superscript 12 in the left hand side of 
Eq.~(\ref{R_12}) means that both the short and the long branches are inside the pore,
and superscript 2 in the left hand side of 
Eq.~(\ref{R_22}) means that only the long branch is inside the pore.
$\tilde{\mathcal{U}}_{\textrm{b}}^{\textrm{I,J}}$ and
$\tilde{\mathcal{V}}_{\textrm{2}}^{\textrm{I,J}}$ are functions of $\nu$, $A_{\nu}$, $\tilde{R}_2$, $\tilde{f}$,
and $\tilde{\phi}$, and their explicit forms can be found in Appendix A.

To have the full solution of the IFTP model when both branches are passing through the pore
Eqs.~(\ref{BD_force}), (\ref{monomer_flux_12}) and (\ref{R_12}) should be solved self-consistently,
and when only the long branch remains one must solve Eqs.~(\ref{BD_force}), 
(\ref{monomer_flux_22}) and (\ref{R_22}).


\section{Molecular dynamics simulations} \label{MD}

To examine the validity of the theory we have performed extensive molecular dynamics (MD) simulations
of a folded polymer pulled through a pore by the monomer separating the two folded branches with $N_{01} \le N_{02}$ monomers. 
We have employed Langevin dynamics simulations using the LAMMPS package 
\cite{Plimpton1995}. The polymer chain is modeled as a coarse-grained self-avoiding bead-spring chain, 
with each bead representing a monomer. The successive beads are connected by the finitely extensible 
nonlinear elastic (FENE) spring interaction given by : 
\begin{equation}
U_{\textrm{FENE}}(r)=-\frac{kR_0^2}{2}\ln\left(1-\frac{r^2}{R_0^2}\right)
\label{FENE_equation}
\end{equation}
where $k$ is the spring constant and $R_0$ is the maximum bond length.
The excluded volume interaction between any two beads, and between a polymer bead and the pore particles 
is given by a repulsive Lenard-Johns (LJ) interaction
\begin{equation}
\begin{split}
U_{\textrm{LJ}}(r)=4\epsilon \left[\left(\frac{\sigma}{r}\right)^{12}- \left(\frac{\sigma}{r}\right)^{6}\right] + \epsilon
\ {\textrm{: if}} \  r \leq r_{\textrm{c}},\\
\  =0 \hspace{+3.45cm} \ {\textrm{: if}} \ r \geq r_{\textrm{c}},
\label{LJ_equation}
\end{split}
\end{equation}
where $r$ is the distance between two beads, $\epsilon$ is the interaction strength and $\sigma$ is the diameter of 
each bead. The cut-off radius is $r_{\textrm{c}}=2^{1/6}\sigma$. The Langevin equation for each $i^{\textrm{th}}$ 
particle of the system is solved as
\begin{equation}
m\ddot{r}_{i}=-\nabla(U_{\textrm{LJ}}+U_{\textrm{FENE}})+ f +F_i^F+F_i^R
\end{equation}
where each polymer bead experiences conservative, frictional and random forces. The frictional force 
$F_i^F = - \eta v_i$, $v_i$ is the monomer velocity, $\eta$ is the solvent friction coefficient, $f$ is the external
pulling force acting on the monomer being pulled, and
$F_i^R$ is the random force with zero mean $\langle F_i^R(t)\rangle=0$ which satisfies the fluctuation-dissipation 
theorem $\langle F_i^R (t) F_j^R(t^\prime)\rangle=6\eta k_{\rm B} T \delta_{ij}\delta(t-t^\prime)$. 
The parameters of our MD simulations in LJ units have been chosen as 
$\sigma=1, \epsilon=1, R_0=1.5\sigma, k=30$ and $\eta=0.7$. Here the external driving force is chosen as $f = 100$,
and $k_{\rm B} T= 1.2$. 
In our model, the mass of each bead $m$ is about 936 amu, its size $\sigma$ corresponds approximately to the Kuhn length 
of a single-stranded DNA, and the interaction strength $\epsilon$ is $3.39\times10^{21}$J at room temperature ($T = 295$ K). 
In LJ units, the time and force scales are 32.1 ps and 2.3 pN, respectively \cite{ikonen2012a,jalal2014}. 
The time step for the integration of the Langevin equation has been chosen as $\textrm{d} t =0.005$ (in LJ units)
during the equilibration of the system and $0.0005$ for the actual translocation process.
It should be mentioned that the width of the pore is small and only one monomer for a linear chain
or two monomers for a folded polymer can be inside the nanopore.

As schematically shown in Fig.~\ref{fig_schematic}, we consider a chain of an odd number of beads, $N_0 =101$
with the pulled bead connecting the two branches placed at the pore. 
Here we consider folded chains with three different branch lengths of 51:51, 31:71 and 11:91. 
At the beginning of the simulations the folded polymer is carefully equilibrated with the pulled bead fixed at the pore. 
Then the constraint is removed and the external pulling
force $f$ starts acting on the bead from {\it cis} towards the {\it trans} side.
As mentioned in the theory section the pulling force is strong enough such that the chain is essentially straightened on 
the {\it trans} side. Translocation time is recorded separately for the short and long segments. Our MD data here
have been averaged over $400-500$ independent runs.

%
\begin{figure*}[t]    
	\begin{center}
        \includegraphics[width=0.99\linewidth]{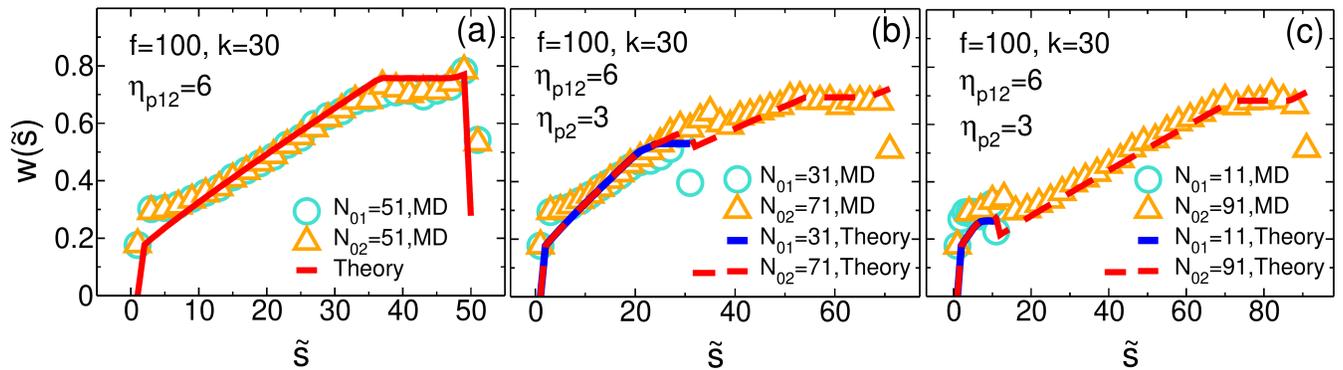}
    \end{center}
\caption{(a) The waiting time distribution $w (\tilde{s})$ as a function of the translocation coordinate $\tilde{s}$ 
for the folded linear polymer chain with equal branches, i.e. $N_{01} = N_{02} = 50 + 1 $, with $N_0 = N_{01} + N_{02} = 101$, 
pore friction $\eta_{\rm p12} =6$, external driving force $f = 100$ and spring constant $k=30$ in the bead-spring model 
used in the MD simulations.
Open light blue circles and open orange triangles are MD data while the solid red line represents the IFTP theory results.
Panels (b) and (c) are the same as panel (a) but for different values of the contour lengths of the two branches
$N_{01} = 31 $ and $N_{02} = 71 $, and $N_{01} = 11 $ and $N_{02} = 91 $, respectively, with constant $N_0 = 101$.
In panels (b) and (c) when both branches are traversing the pore, the pore friction in the theory is $\eta_{\rm p12} =6$,
and after the translocation of the shorter branch to the {\it cis} side it reduces to the fixed value of $\eta_{\rm p 2} =3$.
Open light blue circles and open orange triangles are MD data for short and long branches, respectively,
while the solid blue and the dashed red lines represent the IFTP theory results for the short and long branches, respectively.
}
\label{fig_WT}
\end{figure*}
\begin{figure}[ht]
    \begin{center}
        \includegraphics[width=1.0\linewidth]{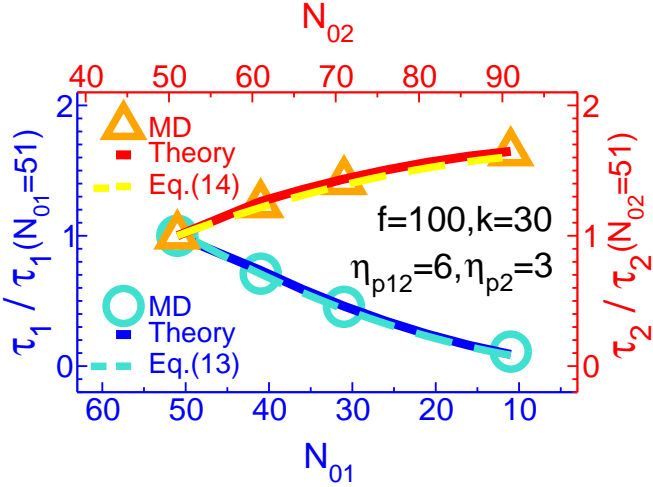}
    \end{center}
\caption{Normalized translocation time $\tau_1 / \tau_1 (N_{01} = 51)$ as a function of 
the contour length for the short branch $N_{01}$, 
and $\tau_2 / \tau_2 (N_{02} = 51)$ as a function of the contour length for the long branch $N_{02}$. 
For the short branch the light blue open circles present the MD data, and the solid blue line 
shows the IFTP theory results. For the long branch the open orange triangles and the solid red 
line present the MD and IFTP theory results, respectively. The data for the short branch 
must be read from the bottom horizontal and left vertical blue axes, while for the long branch 
the data should be read from the top horizontal and right vertical red axes.
The results of the Eqs.~(\ref{tau1}) and (\ref{tau2}) 
are shown in light blue dashed and yellow dashed lines for the short and long branches, respectively.
} 
\label{fig_translocation-time}
\end{figure}
\begin{figure*}[t]    
	\begin{center}
        \includegraphics[width=0.99\linewidth]{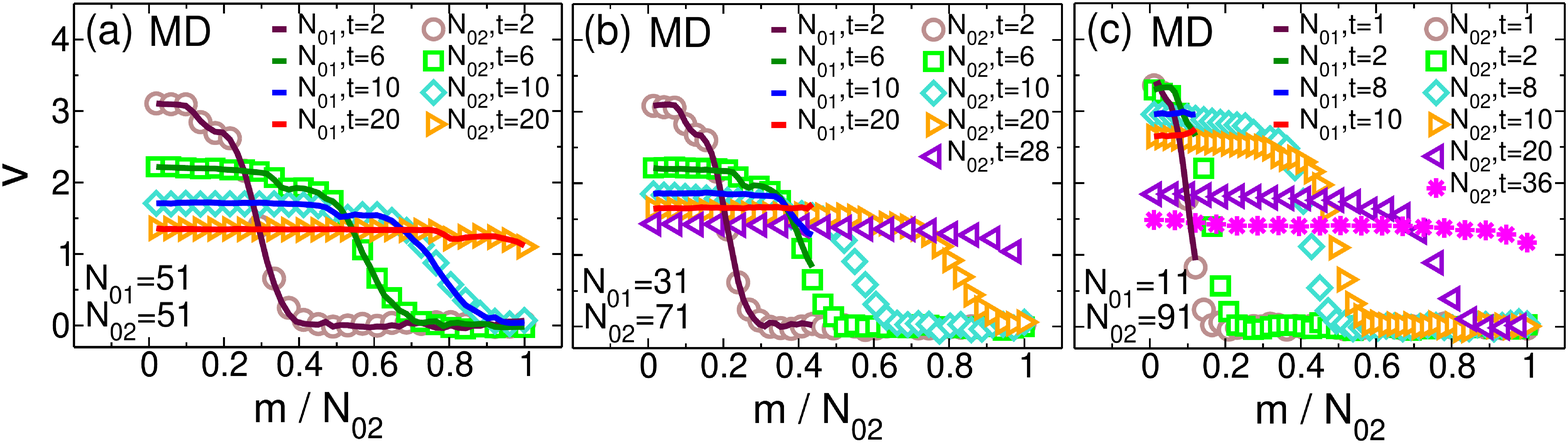}
    \end{center}
\caption{(a) The velocities of the individual monomers from MD simulations as a function of the normalized monomer index, 
$m / N_{02}$, at different moments $t=2-20$ for the symmetric folded chain with $N_{01} =  N_{02} = 51$.
Open symbols show the velocities for the branch 2, while the solid lines present the monomer velocities of 
the branch 1. Panels (b) and (c) are the same as panel (a) but for asymmetrical folded polymer chain with 
branch contour lengths $N_{01} = 31$ and $N_{02} = 71$, and $N_{01} = 11$ and $N_{02} = 91$, respectively. 
In panels (b) and (c) the label of the horizontal axis has been normalized to the contour length of 
the longer branch $N_{02}$, and the data is presented during $t=2-28$ and $t=1-36$, respectively. All data are from MD.
}
\label{fig_mon_vel}
\end{figure*}
%


\section{Results} \label{results}


\subsection{Waiting time distribution} \label{WT}

In order to examine the validity of the IFTP theory we first compare dynamics of 
the translocation process at the monomer level between IFTP theory and MD simulations using the waiting time distribution (WT), which is the 
average time that each bead spends in the pore during 
the process of the translocation.
In Fig.~\ref{fig_WT} we show our data for the three sets of folded polymers. In panel (a) the WT $w (\tilde{s})$ 
has been plotted as a function of the translocation coordinate $\tilde{s}$ for the case wherein the branches have 
the same contour length, i.e. $N_{01} = N_{02} = 51 $, with the total contour length of linear polymer
$N_0 = N_{01} + N_{02} = 101$. 
Here the bead that connects the two branches 
naturally belongs to both of them.
The pore friction used in the IFTP theory is $\eta_{\rm p12} =6$, external driving force which acts on 
the connecting bead is $f = 100$ (both in the theory as well as in the MD simulations) 
and $k=30$ is the spring constant in the bead-spring model used in the MD simulations.
Open light blue circles and open orange triangles are MD data while the solid red line represents the IFTP 
theory results.
Panels (b) and (c) are the same as (a) but for different values of the contour lengths of the two branches
$N_{01} = 31 $ and $N_{02} = 71 $, and $N_{01} = 11 $ and $N_{02} = 91 $, respectively, with constant $N_0 = 101$.
In panels (b) and (c) when both branches are traversing the pore, the pore friction in the IFTP theory is 
chosen as $\eta_{\rm p12} =6$,
and after the translocation of the short branch, it reduces to the fixed value of $\eta_{\rm p 2} =3$.
Open light blue circles and open orange triangles are MD data for short and long branches, respectively,
while the solid blue and the dashed red lines represent the IFTP theory results for the short and long branches, 
respectively. It is obvious from the figure that the agreement between IFTP theory and MD simulations is very good.



\subsection{Translocation time for each branch} \label{Translocation_time}

The central quantity that describes the global dynamics of the polymer translocation through a nanopore
is the average translocation time $\tilde{\tau}$. To find the IFTP translocation time for the short and long branches, $\tau_1$ and $\tau_2$, 
respectively, Eqs.~(\ref{BD_force}), (\ref{monomer_flux_12}), (\ref{monomer_flux_22}), 
(\ref{R_12}) and (\ref{R_22}) must be solved self-consistently.
To compare with MD, in 
Fig.~\ref{fig_translocation-time} the normalized translocation times for the short branch 
$\tau_1 / \tau_1 (N_{01} = 51)$ has been plotted as a function of the contour length of 
the short branch $N_{01}$ (left and bottom blue axes), 
while the normalized translocation time for the long branch $\tau_2 / \tau_2 (N_{02} = 51)$, 
has been plotted as a function of the contour length of the longer branch $N_{02}$ 
(right and top red axes). 
$\tau_1 (N_{01} = 51) = \tau_2 (N_{02} = 51)$ are the translocation times for the two branches of 
a folded linear polymer with contour length of $N_0 = 101$ when the polymer is folded in the middle.
For the short branch the light blue open circles present the MD data, and the solid blue line 
shows the IFTP theory results. On the other hand the open orange triangles and the solid red 
line present the MD and the IFTP theory results, respectively, for the long branch. 
As the figure shows the data are in excellent agreement as already expected from the WT distributions.


\subsection{Scaling of the translocation time}

To find a scaling form for the translocation time for the short branch Eq.~(\ref{monomer_flux_12}) should be
integrated over $N_{1}$ from zero to $N_{01}$ in the TP1-TP2 stage as depicted in Fig.~\ref{fig_schematic}(a)
followed by the integration of $\tilde{R}_{1}$ from $\tilde{R}_{1} (N_{01})$ to zero in the PP1-TP2 stage 
as shown in Fig.~\ref{fig_schematic}(b). This yields the translocation time for the short branch in the SS regime
for the {\it cis} and {\it trans} sides as
\begin{equation}
\tilde{\tau}_1 = \frac{1}{\tilde{f}/2} \bigg( \frac{A_{\nu}}{1+\nu} ~ N_{01}^{1+\nu} + \frac{N_{01}^2}{2} 
+ \tilde{\eta}_{\textrm{p}12} ~ N_{01} \bigg),
\label{tau1}
\end{equation}
where the first, second and the third terms in the r.h.s of Eq.~(\ref{tau1}) are due to 
the mobile sub-branch friction in the {\it cis} side, straightened sub-chain friction in 
the {\it trans} side and the pore friction, respectively. 
Similarly, to find a closed form of the translocation time for the longer branch in the SS regime, 
one needs to integrate 
Eq.~(\ref{monomer_flux_22}) over $N_{2}$ from $N_{01}$ to $N_{02}$ in the $\tau_1$-TP2 stage as presented 
in Fig.~\ref{fig_schematic}(c) followed by the integration of $\tilde{R}_{2}$ from $\tilde{R}_{2} (N_{02})$ 
to zero in the $\tau_1$-PP2 stage as shown in Fig.~\ref{fig_schematic}(d). Following this procedure 
the translocation time for the longer branch in the SS regime for the {\it cis} and {\it trans} sides 
can be written as
\begin{eqnarray}
\hspace{-0.5cm} \tilde{\tau}_2 = \frac{1}{\tilde{f}} \bigg[  \hspace{-0.3cm}
& & + \frac{A_{\nu}}{1+\nu} ~ N_{01}^{1+\nu} + \frac{A_{\nu}}{1+\nu} ~ N_{02}^{1+\nu}
+ \frac{N_{02}^2}{2}-  \frac{N_{01}^2}{2}   \nonumber\\
\hspace{-0.5cm} & &  + 2 \tilde{\eta}_{\textrm{p}12} ~ N_{01}
+ \tilde{\eta}_{\textrm{p}2} ~ ( N_{02} - N_{01} ) + N_{01} N_{02} \bigg],
\label{tau2}
\end{eqnarray}
revealing how the contour lengths of the short and long branches are coupled to each other.
In Fig.~\ref{fig_translocation-time} the results from the Eqs.~(\ref{tau1}) and (\ref{tau2}) 
are shown in light blue and yellow dashed lines for the short and long branches, respectively.
As can be seen there is a very good agreement between the normalized scaling formula, the MD results and the IFTP theory,
although the scaling formulae in Eqs.~(\ref{tau1}) and (\ref{tau2}) have been obtained in the limit of long branches 
in the SS regime.


\subsection{Monomer velocities} \label{MonVel}

\begin{figure*}[t]    
	\begin{center}
        \includegraphics[width=0.99\linewidth]{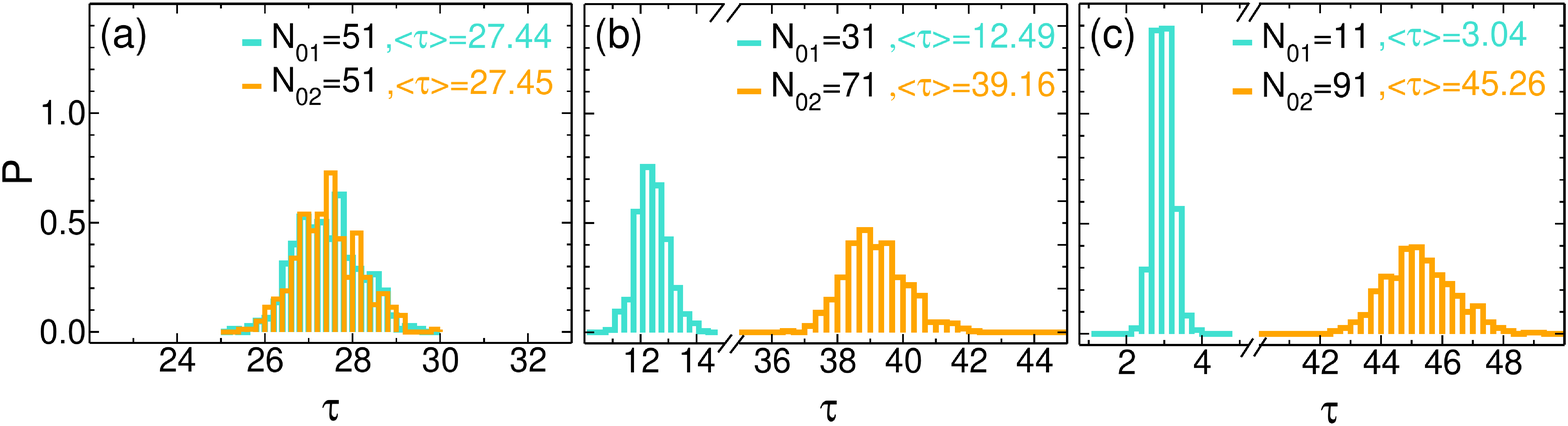}
    \end{center}
\caption{(a) The probability distribution of the translocation time, $P (\tau)$, for both branches of a folded polymer chain 
with total contour length of $N_{0} = 101$, as a function of the translocation time $\tau$ when the folding is symmetric 
and the contour length of both branches are the same as $N_{01} = N_{02} = 51$. Panels (b) and (c) are the same 
as panel (a) but for asymmetrical folded polymer with different sets for the contour lengths of the branches 
$N_{01} = 31$ and $N_{02} = 71$, and $N_{01} =11$ and $N_{02} = 91$, respectively. All data are from MD.
}
\label{fig_time-distribution}
\end{figure*}
In this subsection we present additional data for the monomer velocities from the MD simulations (averaged over $400-500$ successful translocation runs).
In Fig.~\ref{fig_mon_vel}(a) the velocities of the individual monomers for the polymer branches, $v(\textrm{m})$, 
have been plotted as a function of the normalized monomer index, ${m} / N_{02}$, at different moments 
of the translocation process $t=2-20$ for the symmetric folded chain with $N_{01} =  N_{02} = 51$.
The solid lines present the monomer velocities of the branch 1, while open symbols show the velocities for the branch 2. 
Panels (b) and (c) are the same as panel (a) but for an asymmetric folded polymer chain with 
branch contour lengths $N_{01} = 31$ and $N_{02} = 71$, and $N_{01} = 11$ and $N_{02} = 91$, respectively. 
In panel (b) the data is presented during $t=2-28$ and in panel (c) during $t=1-36$.
Here, the individual monomer velocities have been averaged in the direction of the driving force, i.e. horizontal direction from 
the {\it cis} to the {\it trans} side.
As can be seen in panel (a) for the symmetric 51:51 folded chain, the monomer velocities for both branches 
are the same, and the tension propagation occurs in an identical manner for both branches.
At each moment, the velocities of the monomers for both branches that have already moved to the {\it trans} side 
have the same and constant value due to the strong pulling force. For the monomer beads which are in the
{\it cis} side, there is a drop in velocity during the TP1-TP2 stage along both the chain branches and 
the velocity is zero for the non-mobile equilibrium part of the branches. 
In panels (b) and (c) as time passes, the TP1 of the short branch $N_{01}$ ends and the PP1 stage starts.
For example in panel (b) the interval $5<t<15$ is the transient window from TP1 to PP1 stage for the short branch, 
while the long one still is in its TP2 stage and the tension is still propagating along its backbone on the 
{\it cis} side. 
In the PP2 stage, as time progresses, the monomers' velocity of the long branch on the \textit{cis} side sub-branch 
increases and finally becomes equal to that of the {\it trans} side sub-branch. 
Moreover, in all three panels (a), (b) and (c) the velocities of the individual monomers of both branches coincide with each other.
This tells us that the equal monomer flux assumption in the IFTP theory is correct.


\subsection{Translocation time distribution} \label{TT}

Finally, the probability distributions of the translocation time, $P (\tau)$, for the folded polymer with various 
branched contour lengths
are investigated using MD simulations. In panel (a) of Fig.~\ref{fig_time-distribution} the probability distribution of 
the translocation time for both branches of the folded polymer chain with contour length of $N_{0} = 101$, 
has been plotted as a function of the translocation time $\tau$ when the folding is symmetric 
and the contour lengths of both branches are the same as $N_{01} = N_{02} = 51$. Panels (b) and (c) are the same 
as panel (a) but for different sets for the contour lengths of the beaches $N_{01} = 31$ and $N_{02} = 71$, 
and $N_{01} =11$ and $N_{02} = 91$, respectively. 
When the folded chain becomes more asymmetrical, i.e. the contour length of one branch gets shorter
and in contrast the other one becomes longer, the probability distributions for different branches 
are separated more from each other. As can be seen in Fig.~\ref{fig_time-distribution}(c) the separation 
is more pronounced than panel (b). 
Moreover, for shorter branch the width of the probability distribution is narrower due 
to the decrease in the spatial fluctuations of the branch configurations.


\section{Summary and conclusion} \label{conclusions}

In this work, we have theoretically and computationally studied the translocation dynamics of a singly-folded polymer chain pulled through a 
nanopore by applying a pulling force on the monomer connecting the two branches. The pulling force initiates a tension front that propagates
though the folds during translocation. To properly treat this, we have generalized the IFTP theory to the present case.  
We have also performed extensive MD simulations of a coarse-grained bead-spring model to benchmark the theory.
The WT distribution obtained from the 
IFTP theory has been compared with MD one showing good agreement at the monomer level dynamics.
Then, the global dynamics of the translocation has been examined by looking at the translocation time for each branch
obtained from the IFTP theory and MD simulations. Again the results of the IFTP theory are in excellent agreement 
with the MD simulations. We have also analytically derived scaling forms for the average translocation time
from the IFTP theory showing explicitly how the two branches are dynamically coupled. 
For both branches the effective translocation exponent, $\alpha$, which is defined as 
$\tau \sim N_{01}^{\alpha}$ is between 1 and 2. While $\alpha$ for the short branch depends on $N_{01}$ 
(see Eq.~(\ref{tau1})), for the long branch the translocation exponent depends on the values of both 
contour lengths $N_{01}$ and $N_{02}$ (see Eq.~(\ref{tau2})).

Finally, we have used MD simulations to characterise the
velocities of the individual monomers for both branches, as well as the probability distribution 
of the translocation time. As the system becomes more asymmetrical, i.e. one branch gets shorter 
while the other one becomes longer, the probability distributions of the translocation time distributions for 
the short and long branches separate. Moreover, the width of the distribution becomes narrower
for the short branch.


\section{Appendix A} \label{Appendix}

In this Appendix the explicit forms of $\tilde{\mathcal{U}}_{\textrm{b}}^{\textrm{I}}$ and
$\tilde{\mathcal{V}}_{\textrm{2}}^{\textrm{I}}$ as functions of $\nu$, $A_{\nu}$, $\tilde{R}_2$, $\tilde{f}$,
and $\tilde{\phi}$ are written. 
In the TP1-TP2 stage (Fig.~\ref{fig_schematic}(a)) the equations of motion of the location of the tension 
fronts for branch 1 and branch 2 are similar to each other as 
\begin{equation}
\dot{\tilde{R}}_{\textrm{1}}^{\textrm{I,J},12} = \tilde{\mathcal{U}}_{\textrm{1}}^{\textrm{I,J}} ; \hspace{+1.0cm}
\dot{\tilde{R}}_{\textrm{2}}^{\textrm{I,J},12} = \tilde{\mathcal{U}}_{\textrm{2}}^{\textrm{I,J}} ,
\label{R_12_App}
\end{equation}
where $\textrm{I}$ stands for different regimes of SSC, SFC and TRC, $\textrm{J}$ denotes different stages of TP1 and TP2, 
superscript 12 in the left hand side of 
Eq.~(\ref{R_12_App}) means that both the short as well as the long branches are inside the pore,
\begin{eqnarray}
\mathcal{U}_{\textrm{1}}^{\textrm{SSC1,TP1}} &=& \mathcal{U}_{\textrm{2}}^{\textrm{SSC2,TP2}} 
= \frac{ B (\tilde{R}_1) \tilde{\phi}_1 }{ 1 - B (\tilde{R}_1)  }  ; \nonumber\\
\mathcal{U}_{\textrm{1}}^{\textrm{SFC1,TP1}} &=& \mathcal{U}_{\textrm{2}}^{\textrm{SFC2,TP2}} 
= \frac{ - B (\tilde{R}_1) \tilde{\phi}_1^2 \mathcal{L}_{\textrm{SFC1}} }
{ 1 + B (\tilde{R}_1) \tilde{\phi}_1 \mathcal{L}_{\textrm{SFC1}} }  ; \nonumber\\
\mathcal{U}_{\textrm{1}}^{\textrm{TRC1,TP1}} &=& \mathcal{U}_{\textrm{2}}^{\textrm{TRC2,TP2}} = \nonumber\\
&& \hspace{-2.5cm}= \frac{ B (\tilde{R}_1) \big[ - \tilde{\phi}_1^2 \mathcal{L}_{\textrm{TRC1}} + 
\tilde{\phi}_1 - \tilde{\phi}_1 (\tilde{\phi}_1 \tilde{R}_1)^{(\nu-1)/\nu} \big] }
{ 1 + B (\tilde{R}_1)  \tilde{\phi}_1 \mathcal{L}_{\textrm{TRC1}} }  , 
\label{U_12_SSC_TP}
\end{eqnarray}
with
\begin{eqnarray}
\hspace{-0.5cm}  B (\tilde{R}_1) &=& \nu A_{\nu}^{1/\nu} \tilde{R}_1^{(\nu -1)/\nu}; \nonumber\\
\hspace{-0.5cm} \tilde{\phi}_1 (\tilde{t}) &=& \tilde{\phi}_2 (\tilde{t}) = \frac{ \tilde{f} / 2 }
{ \tilde{R}_1 + \tilde{\eta}_{\textrm{p12}} + \tilde{s}_1 }, \nonumber\\
\hspace{-0.5cm} \mathcal{L}_{\textrm{SFC1}} &=&
\frac{\nu -1}{(2\nu-1) \big[ \tilde{R}_1 + \tilde{s}_1 + \tilde{\eta}_{\textrm{p12}} \big]  \tilde{\phi}_1^2 } 
- \frac{1}{\tilde{\phi}_1} ; \nonumber\\
\hspace{-0.5cm}  \mathcal{L}_{\textrm{TRC1}} &=& \frac{\tilde{\phi}_1^{-(1+\nu)/\nu} \tilde{R}_1^{(\nu-1)/\nu} }
{ \tilde{R}_1 + \tilde{s}_1 + \tilde{\eta}_{\textrm{p12}} } 
\big[ \frac{\nu-1}{2\nu-1}  \tilde{\phi}_1 \tilde{R}_1 - \tilde{f} /2 \big] .
\label{U_12_SFC_TP}
\end{eqnarray}
In the PP1-TP2 stage (Fig.~\ref{fig_schematic}(b)) the equations of motion for the tension front location 
for branch 1 and branch 2, in the SSC1-SSC2 regime are
written as
\begin{equation}
\hspace{-0.5cm} \dot{\tilde{R}}_{\textrm{1}}^{\textrm{SSC1,PP1},12} = - \tilde{\phi}_1 ; \hspace{+0.5cm}
\dot{\tilde{R}}_{\textrm{2}}^{\textrm{SSC2,TP2},12} =  \frac{ B (\tilde{R}_2) \tilde{\phi}_2 } { 1- B (\tilde{R}_2) },
\label{R_12_SSC1-PP1_SSC2-TP2}
\end{equation}
where 
\begin{equation}
\tilde{\phi}_1 = \tilde{\phi}_2 = \tilde{\phi} =
\frac{ \tilde{f} } { \tilde{R}_1 + \tilde{R}_2 + \tilde{s}_1 + \tilde{s}_2 + 2 \tilde{\eta}_{\textrm{p12}} } ,
\label{phi_12}
\end{equation}
is the monomer flux when both branches are inside the nanopore.

For the PP1-TP2 stage (Fig.~\ref{fig_schematic}(b)) in the SSC1-SFC2 regime the equations of motion 
for $\tilde{R}_1$ and $\tilde{R}_2$ are coupled to each other as 
\begin{equation}
\dot{\tilde{R}}_{\textrm{1}}^{\textrm{SSC1,PP1},12} = - \tilde{\phi} ;  \hspace{+0.15cm}
\dot{\tilde{R}}_{\textrm{1}}^{\textrm{SSC1,PP1},12} + G_2 \dot{\tilde{R}}_{\textrm{2}}^{\textrm{SFC2,TP2},12} =  H_2,
\label{R_12_SSC1-PP1_SFC2-TP2}
\end{equation}
in the SSC1-TRC2 regime they are 
\begin{equation}
\hspace{-0.04cm} \dot{\tilde{R}}_{\textrm{1}}^{\textrm{SSC1,PP1},12} = - \tilde{\phi} ; \hspace{+0.15cm}
\hspace{-0.04cm} \dot{\tilde{R}}_{\textrm{1}}^{\textrm{SSC1,PP1},12} + \mathbb{G}_2 \dot{\tilde{R}}_{\textrm{2}}^{\textrm{TRC2,TP2},12} = \mathbb{H}_2,
\label{R_12_SSC1-PP1_TRC2-TP2}
\end{equation}
in the SFC1-SSC2 regime
\begin{eqnarray}
\hspace{-0.5cm} \dot{\tilde{R}}_{\textrm{1}}^{\textrm{SFC1,PP1},12} 
+ \dot{\tilde{R}}_{\textrm{2}}^{\textrm{SSC2,TP2},12} \frac{\mathcal{F}-1}{\mathcal{F}} &=& 
\frac{ (1-2\mathcal{F}) \tilde{\phi} }{\mathcal{F}} ;  \nonumber\\
\hspace{-0.5cm} \dot{\tilde{R}}_{\textrm{2}}^{\textrm{SSC2,TP2},12} &=& 
\frac{ B ( \tilde{R}_2 ) \tilde{\phi} }{ 1 - B (\tilde{R}_2) }   ,
\label{R_12_SFC1-PP1_SSC2-TP2}
\end{eqnarray}
in the SFC1-SFC2 regime
\begin{eqnarray}
\dot{\tilde{R}}_{\textrm{1}}^{\textrm{SFC1,PP1},12} 
+ \dot{\tilde{R}}_{\textrm{2}}^{\textrm{SFC2,TP2},12} \frac{\mathcal{F}-1}{\mathcal{F}} &=& 
\frac{ (1-2\mathcal{F}) \tilde{\phi} }{\mathcal{F}};  \nonumber\\
\dot{\tilde{R}}_{\textrm{1}}^{\textrm{SFC1,PP1},12} + \dot{\tilde{R}}_{\textrm{2}}^{\textrm{SFC2,TP2},12} G_2
&=& H_2  ,
\label{R_12_SFC1-PP1_SFC2-TP2}
\end{eqnarray}
in the SFC1-TRC2 regime
\begin{eqnarray}
\dot{\tilde{R}}_{\textrm{1}}^{\textrm{SFC1,PP1},12} 
+ \dot{\tilde{R}}_{\textrm{2}}^{\textrm{TRC2,TP2},12} \frac{\mathcal{F}-1}{\mathcal{F}} &=& 
\frac{ (1-2\mathcal{F}) \tilde{\phi} }{\mathcal{F}} ;  \nonumber\\
\dot{\tilde{R}}_{\textrm{1}}^{\textrm{SFC1,PP1},12} + 
\dot{\tilde{R}}_{\textrm{2}}^{\textrm{TRC2,TP2},12} \mathbb{G}_2
&=& \mathbb{H}_2  ,
\label{R_12_SFC1-PP1_TRC2-TP2}
\end{eqnarray}
in the TRC1-SSC2 regime
\begin{eqnarray}
\dot{\tilde{R}}_{\textrm{1}}^{\textrm{TRC1,PP1},12} + 
\dot{\tilde{R}}_{\textrm{2}}^{\textrm{SSC2,TP2},12} G_1
&=& H_1  ;  \nonumber\\
\dot{\tilde{R}}_{\textrm{2}}^{\textrm{SSC2,TP2},12} &=& 
\frac{ B(\tilde{R}_2) \tilde{\phi} }{ 1 - B(\tilde{R}_2) } ,
\label{R_12_TRC1-PP1_SSC2-TP2}
\end{eqnarray}
in the TRC1-SFC2 regime
\begin{eqnarray}
\dot{\tilde{R}}_{\textrm{1}}^{\textrm{TRC1,PP1},12} + 
\dot{\tilde{R}}_{\textrm{2}}^{\textrm{SFC2,TP2},12} G_1
&=& H_1  ;  \nonumber\\
\dot{\tilde{R}}_{\textrm{1}}^{\textrm{TRC1,PP1},12} + 
\dot{\tilde{R}}_{\textrm{2}}^{\textrm{SFC2,TP2},12} G_2
&=& H_2  ,
\label{R_12_TRC1-PP1_SFC2-TP2}
\end{eqnarray}
and in the TRC1-TRC2 regime
\begin{eqnarray}
\dot{\tilde{R}}_{\textrm{1}}^{\textrm{TRC1,PP1},12} + 
\dot{\tilde{R}}_{\textrm{2}}^{\textrm{TRC2,TP2},12} G_1
&=& H_1  ;  \nonumber\\
\dot{\tilde{R}}_{\textrm{1}}^{\textrm{TRC1,PP1},12} + 
\dot{\tilde{R}}_{\textrm{2}}^{\textrm{TRC2,TP2},12} \mathbb{G}_2
&=& \mathbb{H}_2  ,
\label{R_12_TRC1-PP1_TRC2-TP2}
\end{eqnarray}
where
\begin{eqnarray}
G_{\textrm{b}} &=& 
\frac{ B (\tilde{R}_{\textrm{b}}) \mathcal{F} - 1 }{ B (\tilde{R}_{\textrm{b}}) (\mathcal{F}-1) } ; \nonumber\\
H_{\textrm{b}} &=& 
\frac{ B (\tilde{R}_{\textrm{b}}) (1 - 2 \mathcal{F}) \tilde{\phi} }{ B (\tilde{R}_{\textrm{b}}) (\mathcal{F}-1) } ; \nonumber\\
\mathbb{G}_2 &=& \frac{ 1 + (\mathcal{F}-1) \tilde{\phi} \tilde{R}_2 - B^{-1} (\tilde{R}_2)~ (\tilde{\phi} \tilde{R}_2)^{(1-\nu)/\nu} }
{ (\mathcal{F}-1) \tilde{\phi} \tilde{R}_2 }; \nonumber\\
\mathbb{H}_2 &=& \frac{ 2 \tilde{\phi} (1 - \mathcal{F}) \tilde{\phi} \tilde{R}_2 -\tilde{\phi} (\tilde{\phi} \tilde{R}_2)^{(1-\nu)/\nu} }
{ (\mathcal{F}-1) \tilde{\phi} \tilde{R}_2 }; \nonumber\\
\mathcal{F}   &=& 1+ \frac{1-\nu}{2\nu-1} \frac{1}{\tilde{f}} .
\label{U_12_SFC_TP}
\end{eqnarray}

In the $\tau_1$-TP2 stage for the SSC2, SFC2 and TRC2 regimes (Fig.~\ref{fig_schematic}(c)), 
wherein the translocation process for the shorter branch has been completed and only the longer branch is
inside the nanopore, the equations of motion for $\tilde{R}_2$ are obtained as
\begin{eqnarray}
\hspace{-1.0cm} \dot{\tilde{R}}_2^{\textrm{SSC2,TP2},2}  &=& 
\frac{ B (\tilde{R}_2 ) \tilde{\phi}_2 }{ 1 - B (\tilde{R}_2 )}; \nonumber\\ 
\hspace{-1.0cm} \dot{\tilde{R}}_2^{\textrm{SFC2,TP2},2} &=& 
\frac{ B (\tilde{R}_2 ) \tilde{\phi}_2 \mathcal{F} }{ 1 - B (\tilde{R}_2 ) \mathcal{F} }; \nonumber\\
\hspace{-1.0cm} \dot{\tilde{R}}_2^{\textrm{TRC2,TP2},2}  &=&  \nonumber\\
&& \hspace{-2.0cm} = \frac{ B (\tilde{R}_2 )\big[ - \tilde{\phi}_2 (\tilde{\phi}_2 \tilde{R}_2)^{(2\nu-1)/\nu} (1-\mathcal{F})
+ \tilde{\phi} \big] }{ 1 + B (\tilde{R}_2 ) (\tilde{\phi}_2 \tilde{R}_2)^{(\nu-1)/\nu} 
\big[ \tilde{\phi}_2 \tilde{R}_2 (1-\mathcal{F}) -1  \big]} ,
\label{R_12_SFC1-PP1_TRC2-TP2}
\end{eqnarray}
while in the $\tau_1$-PP2 stage (Fig.~\ref{fig_schematic}(d)) the equations of motion 
for $\tilde{R}_2$ are
\begin{eqnarray}
\hspace{-0.5cm} \dot{\tilde{R}}_2^{\textrm{SSC2,PP2},2} &=& 
- \tilde{\phi}_2 ; \nonumber\\ 
\hspace{-0.5cm} \dot{\tilde{R}}_2^{\textrm{SFC2,PP2},2} &=& 
- \tilde{\phi}_2 ; \nonumber\\
\hspace{-0.5cm} \dot{\tilde{R}}_2^{\textrm{TRC2,PP2},2} &=& 
\frac{ - (\tilde{\phi}_2 \tilde{R}_2)^{(2\nu-1)/\nu} (1-\mathcal{F})
+ \tilde{\phi} }{ (\tilde{\phi}_2 \tilde{R}_2)^{(\nu-1)/\nu} 
\big[ \tilde{\phi}_2 \tilde{R}_2 (1-\mathcal{F}) -1  \big]} .
\label{R_12_SFC1-PP1_TRC2-TP2}
\end{eqnarray}
In the $\tau_1$-TP2 and $\tau_1$-PP2 stages the monomer flux is
\begin{equation}
\tilde{\phi}_2 = 
\frac{ \tilde{f} } { \tilde{R}_2 + \tilde{s}_2 + N_{01} + \tilde{\eta}_{\textrm{p2}} } .
\label{phi_12}
\end{equation}

Finally, it should be mentioned that in order to obtain the time evolution of the tension fronts, 
in the left hand side of the above equations of motion the following replacement must be done
$\dot{\tilde{R}}_{1}^{\textrm{I,J},12} = \dot{\tilde{R}}_1$, 
$\dot{\tilde{R}}_{2}^{\textrm{I,J},12} = \dot{\tilde{R}}_2$. 
In the equations $\tilde{R}_1$ and $\tilde{R}_2$
present the time evolution of the tension front in the corresponding regimes and stages 
mentioned by superscripts.


\begin{acknowledgments}
S.C. and B.G. acknowledge DAE-BRNS (37(2)/ 14/08/2016-BRNS/37022) for computer facilities. 
B.G. acknowledges DAE-BRNS (37(2)/ 14/08/2016-BRNS/37022) and IISER Pune for fellowship.
T.A-N. has been supported in part by the Academy of Finland through its 
PolyDyna (no. 307806) and QFT Center of Excellence Program grants (no. 312298).
J.S. thanks Takahiro Sakaue for enlightening discussions.
\end{acknowledgments}


\end{document}